\documentstyle[amssymb,theorem]{article}
\theoremstyle{plain} 
\newtheorem{theorem}{Theorem}[section]
\theoremstyle{plain} 

\theoremstyle{plain} 
\newtheorem{lem}[theorem]{Lemma}
\theoremstyle{plain} 

{\theorembodyfont{\upshape}\theoremstyle{break}
\newtheorem{rem}[theorem]{Remark}}
\newcommand{\qed}{\hfill\hbox{\rule[-2pt]{3pt}{6pt}}}
\newcommand{\comment}[1]{}
\begin{document}
\begin{center}
{\Large
Determinant Formulas for 
the Toda and Discrete Toda Equations}
\end{center}
\vskip1cm
\begin{center}
Kenji Kajiwara$^{a}$, 
Tetsu Masuda$^{a}$,
Masatoshi Noumi$^{b}$, \\
Yasuhiro Ohta$^{c}$  and
Yasuhiko Yamada$^{b}$
\end{center}
\vskip1cm
\begin{center}
a: Department of Electrical Engineering, \\
Doshisha University, Kyotanabe, Kyoto 610-0321, Japan\\
b: Department of Mathematics, \\
Kobe University, Rokko, Kobe 657-8501, Japan\\
c: Department of Applied Mathematics\\
Hiroshima University, Higashi-Hiroshima 739-8527, Japan\\
\end{center}
\vskip1cm
{\bf Abstract: }
Determinant formulas for the general solutions of the Toda 
and discrete Toda equations 
are presented. 
Application to the $\tau$ functions for the 
Painlev\'e equations is also discussed.
\vskip1cm
\section{Introduction}
In the works \cite{KO1, KO2, O1, O2, O3, O4, KM1, NY4}, 
the determinant formulas for the $\tau$ functions
of the Painlev\'e equations are obtained.
These determinant formulas arise as a consequence of the Toda equation
which describes B\"acklund (or Schlesinger) transformations of the
Painlev\'e equations \cite{O1,O2,O3,O4}.
They provide a proof of the miraculous polynomiality 
of the special polynomials arising as the special solutions of
the Painlev\'e equations \cite{Um1, Um2}.

Recently, it is clarified that these determinant formulas can be also 
applied not only for the special (classical) solutions but also for generic 
(transcendental) ones \cite{Um2,KM2}. 
It is natural to expect the existence of such determinant formulas
for the general solutions of the Toda equation independent of the Painlev\'e
equations and their (special) solutions.

In this paper, we will give a formula of Hankel type determinant
for the solution $\tau_n$ of the Toda equation 
(viewed as a recurrence relation)
$$
\tau_n'' \tau_n-({\tau_n}')^2=\tau_{n-1} \tau_{n+1},
$$
for general initial conditions $\tau_0$ and $\tau_1$.
In the case of $\tau_0=1$, such a formula is known as Darboux's
formula (cf. \cite{O1}).
As an application of the formula, we will consider
the $\tau$ functions of the Painlev\'e equations.
We also  present a similar determinant formula for the discrete Toda equation
$$
\rho_n^{l+1}\rho_n^{l-1} - \left(\rho_n^{l}\right)^2=\varepsilon^2
\rho_{n+1}^{l+1}\rho_{n-1}^{l-1}.
$$

\section{Determinant formulas}
In this section, we present the determinant formulas for general
solutions of Toda and discrete Toda equations.

We consider the following recursion relation of a sequence
$\{\tau_n\}_{n\in \mathbb{Z}}$,
\begin{equation}
 \tau_n^{\prime\prime}\tau_n - \tau_n^{\prime 2}=\tau_{n+1}\tau_{n-1}-\psi\varphi\tau_n^2,
\label{Toda:tau}
\end{equation}
with
\begin{equation}
 \tau_{-1}=\psi,\quad \tau_0=1,\quad \tau_1=\varphi,
\label{Toda:ini}
\end{equation}
where $\psi$ and $\varphi$ are arbitrary functions, and ${}^\prime$ denotes
a derivation. Equation (\ref{Toda:tau}) is called the Toda equation in 
bilinear form. For given $\psi$ and $\varphi$, $\tau_n$ are
uniquely determined as rational functions in them and their derivatives.
However, as we shall show below, $\tau_n$ are actually polynomials in
$\psi$, $\varphi$ and their derivatives, and
moreover, they are expressed in determinantal forms.

\begin{theorem}\label{main1}
Let $\{a_n\}_{n\in \mathbb{N}}$,
$\{b_n\}_{n\in\mathbb{N}}$ be two sequences defined recursively as
\begin{equation}
 a_n=a_{n-1}^\prime + \psi\sum_{i+j=n-2\atop i,j\geq 0}a_ia_j,\quad a_0=\varphi,\label{Toda:rec1}
\end{equation}
\begin{equation}
 b_n=b_{n-1}^\prime + \varphi\sum_{i+j=n-2\atop i,j\geq 0}b_ib_j,\quad b_0=\psi.\label{Toda:rec2}
\end{equation}
For any integer $n$, we define $|n|\times |n|$ Hankel determinant $\tau_n$ by
\begin{equation}
 \tau_n=\left\{
  \begin{array}{cc}
\left|\begin{array}{cccc}
 a_0&a_1 &\cdots &a_{n-1} \\
 a_1&a_2 &\cdots &a_{n} \\
\vdots & \vdots &\ddots & \vdots\\
 a_{n-1}&a_n &\cdots &a_{2n-2} 
      \end{array}\right|, &n>0,\\
1, &n=0,\\
\left|\begin{array}{cccc}
 b_0&b_1 &\cdots &b_{-n-1} \\
 b_1&b_2 &\cdots &b_{-n} \\
\vdots & \vdots &\ddots & \vdots\\
 b_{-n-1}&b_{-n} &\cdots &b_{-2n-2} 
      \end{array}\right|, &n<0,\\
\end{array}
\right.\label{Toda:det}
\end{equation}
Then, $\tau_n$ satisfies equation (\ref{Toda:tau}) with initial condition (\ref{Toda:ini}).
\end{theorem}
\begin{rem}
 \begin{enumerate}
  \item Equation (\ref{Toda:tau}) is transformed to the bilinear form of 
	Toda equation in ``usual'' form,
\begin{equation}
  \sigma_n^{\prime\prime}\sigma_n - \sigma_n^{\prime 2}=\sigma_{n+1}\sigma_{n-1},
\label{Toda:sigma}
\end{equation}
by applying suitable gauge transformation on $\tau_n$, where 
$\sigma_n$ is given by
\begin{equation}
 (\log \sigma_n)'' = (\log\tau_n)''+\psi\varphi.
\end{equation}
  \item Since $\sigma_n$ involves two arbitrary functions, Theorem
	\ref{main1} gives a determinant formula for the general solution
	of the Toda equation.
  \item In the case of $\varphi=0$ or $\psi=0$, Theorem \ref{main1}
	recovers the well-known Darboux formula, namely, the determinant expression
	for solutions of the so-called Toda molecule equation\cite{LS,H1}.
 \end{enumerate}
\end{rem}
It is also possible to construct a similar formula for the discrete Toda
equation.  Let $\Phi^l$ and $\Psi^l$ be arbitrary functions in $l$, and
$\{\kappa_n^l\}_{n\in \mathbb{Z}}$ a sequence defined by
\begin{equation}
 \kappa_n^{l+1}\kappa_n^{l-1}-(1-\varepsilon^2\Phi^{l+1}\Psi^{l-1})\left(\kappa_n^l\right)^2
=\varepsilon^2\kappa_{n+1}^{l+1}\kappa_{n-1}^{l-1},\label{dToda:kappa}
\end{equation}
\begin{equation}
 \kappa_{-1}^l=\Phi^l,\quad \kappa_0^l=1,\quad \kappa_1^l=\Psi^l,
\label{dToda:ini}
\end{equation}
where $\varepsilon$ is a paramater corresponding to the lattice interval of $l$.
Equation (\ref{dToda:kappa}) is called the discrete Toda equation in
bilinear form \cite{H2}. Then we have:
\begin{theorem}\label{main2}
Let $\{c_k^l\}_{k\in \mathbb{N}}$,
$\{d_k^l\}_{k\in\mathbb{N}}$ be two sequences defined recursively as
\begin{equation}
c_{k}^l=\frac{c_{k-1}^l-c_{k-1}^{l-1}}{\varepsilon}+\Psi^{l-2}
\sum_{i+j=k-2\atop i,j\geq 0}\left(c_i^l-\varepsilon c_{i+1}^l\right)c_{j}^{l-1},
\quad c_0^l=\Phi^l,\label{dToda:rec1}
\end{equation}
\begin{equation}
d_{k}^l=\frac{d_{k-1}^{l+1}-d_{k-1}^{l}}{\varepsilon}+\Phi^{l+2}
\sum_{i+j=k-2\atop i,j\geq 0}\left(d_i^l+\varepsilon c_{i+1}^l\right)d_{j}^{l+1},\quad
d_0^l=\Psi^l.\label{dToda:rec2}
\end{equation}
For any integer $n$, we define $|n|\times |n|$ Hankel determinant $\tau_n$ by
\begin{equation}
 \kappa^l_n=\left\{
  \begin{array}{cc}
\left|\begin{array}{cccc}
 c^l_0&c^l_1 &\cdots &c^l_{n-1} \\
 c^l_1&c^l_2 &\cdots &c^l_{n} \\
\vdots & \vdots &\ddots & \vdots\\
 c^l_{n-1}&c^l_n &\cdots &c^l_{2n-2} 
      \end{array}\right|, &n>0,\\
1, &n=0,\\
\left|\begin{array}{cccc}
 d^l_0&d^l_1 &\cdots &d^l_{-n-1} \\
 d^l_1&d^l_2 &\cdots &d^l_{-n} \\
\vdots & \vdots &\ddots & \vdots\\
 d^l_{-n-1}&d^l_{-n} &\cdots &d^l_{-2n-2} 
      \end{array}\right|, &n<0,\\
\end{array}
\right.\label{dToda:det}
\end{equation}
Then, $\kappa^l_n$ satisfies equation (\ref{dToda:kappa}) with initial condition (\ref{dToda:ini}).
\end{theorem}
\begin{rem}
 \begin{enumerate}
  \item Equation (\ref{dToda:kappa}) is transformed to the bilinear form of 
discrete Toda equation in ``usual'' form,
\begin{equation}
  \rho_n^{l+1}\rho_n^{l-1} - \left(\rho_n^{l}\right)^2=\varepsilon^2
\rho_{n+1}^{l+1}\rho_{n-1}^{l-1},
\label{dToda:rho}
\end{equation}
by introducing $\rho$ by
\begin{equation}
 \rho_n^l = \frac{1}{\prod_{i=l_0}^l\prod_{j=i_0}^i\left(1-\varepsilon^2\Phi^j\Psi^{j-2}\right)}
\kappa_n^l.
\end{equation}
  \item Since $\rho_n$ involves two arbitrary functions, Theorem
	\ref{main2} gives a determinant expression for general solution
	of the discrete Toda equation.
  \item In the case of $\varphi=0$ or $\psi=0$, Theorem \ref{main2}
	recovers the determinant expression
	for solution of the so-called discrete Toda molecule
	equation\cite{H3}.
  \item Discrete Toda equation (\ref{dToda:kappa}) and 
	its solution (\ref{dToda:rec1})-(\ref{dToda:det})
	reduce to the Toda equation (\ref{Toda:tau}) and 
	its solution (\ref{Toda:rec1})-(\ref{Toda:det}) in
	the limit of $\varepsilon\to 0$, respectively.
 \end{enumerate}
\end{rem}
Let us first prove Theorem \ref{main1}. We consider the case of $n>
0$. 
Let $D$ be the determinant of an $(n+1)\times (n+1)$ matrix $X$, 
and $D\left(\begin{array}{cccc}
i_1&i_2&\cdots&i_k  \\
j_1&j_2&\cdots&j_k\end{array}\right)$ the determinant 
of the matrix obtained from X by removing the rows with indices $i_1,\cdots,i_k$
and the columns with indices $j_1,\cdots, j_k$. 
Then we have well-known Jacobi's formula (Lewis Carroll's formula)
\begin{equation} 
 D\left(\begin{array}{c}
   n\\
   n
	\end{array}\right)
 D\left(\begin{array}{c}
   n+1\\
   n+1
	\end{array}\right)
- D\left(\begin{array}{c}
   n\\
   n+1
	\end{array}\right)
 D\left(\begin{array}{c}
   n+1\\
   n
	\end{array}\right)
=D\cdot  D\left(\begin{array}{cc}
  n& n+1\\
  n& n+1
	\end{array}\right).\label{Toda:Jacobi}
\end{equation}
We have the following differential formula for $\tau_n$:
\begin{lem}\label{Toda:shift}
Putting $D \equiv\tau_{n+1}$, we have,
 \begin{equation}
   D\left(\begin{array}{c}
   n+1\\
   n+1
	\end{array}\right)=\tau_n,\quad
D\left(\begin{array}{cc}
  n& n+1\\
  n& n+1
	\end{array}\right)=\tau_{n-1},\label{Toda:shift1}
 \end{equation}
\begin{equation}
    D\left(\begin{array}{c}
   n\\
   n+1
	\end{array}\right)=
   D\left(\begin{array}{c}
   n+1\\
   n
	\end{array}\right)=\tau_n^\prime,\label{Toda:shift2}
\end{equation}
\begin{equation}
    D\left(\begin{array}{c}
   n\\
   n
	\end{array}\right)=\tau_n^{\prime\prime}+\varphi\psi\tau_n.\label{Toda:shift3}
\end{equation}
\end{lem}
Then, Theorem \ref{main1} follows immediately from equation
(\ref{Toda:Jacobi}) and Lemma \ref{Toda:shift}. Therefore, it suffices
to prove Lemma \ref{Toda:shift}.

\noindent {\it Proof of Lemma \ref{Toda:shift}} Equation
(\ref{Toda:shift1}) is obvious by definition. To show equation
(\ref{Toda:shift2}), we consider the following equality,
\begin{eqnarray}
&&     D\left(\begin{array}{c}
   n\\
   n+1
	\end{array}\right)=
   D\left(\begin{array}{c}
   n+1\\
   n
	\end{array}\right)\nonumber\\
&=&
\left(\begin{array}{cccc}
 a_1&a_2 &\cdots &a_n \\
 a_2&a_3 &\cdots &a_{n+1} \\
\vdots &\vdots &\ddots &\vdots\\
 a_{n}&a_{n+1} &\cdots &a_{2n-1}
      \end{array}\right)\cdot
\left(\begin{array}{cccc}
 \Delta_{11}&\Delta_{12} &\cdots &\Delta_{1n} \\
 \Delta_{21}&\Delta_{22} &\cdots &\Delta_{2n} \\
\vdots &\vdots &\ddots &\vdots\\
 \Delta_{n1}&\Delta_{n2} &\cdots &\Delta_{nn} 
      \end{array}\right).\label{Toda:prod1}
\end{eqnarray}
Here, for $n\times n$ matrices $A=(A_{ij})$ and $B=(B_{ij})$, $A\cdot B$ denotes
\begin{equation}
 A\cdot B=\sum_{i,j=1}^nA_{ij}B_{ij}={\rm Tr}\ A\,{}^t\!B,
\end{equation}
which is the standard scalar product of matrices, and $\Delta_{ij}$ is
an $(i,j)$-cofactor of $\tau_n$. The first matrix of equation
(\ref{Toda:prod1}) is rewritten by using the recursion relation
(\ref{Toda:rec1}) as
\begin{eqnarray}
&& \left(\begin{array}{cccc}
 a_1&a_2 &\cdots &a_n \\
 a_2&a_3 &\cdots &a_{n+1} \\
\vdots &\vdots &\ddots &\vdots\\
 a_{n}&a_{n+1} &\cdots &a_{2n-1}
      \end{array}\right)
=
 \left(\begin{array}{cccc}
 a^\prime_1&a^\prime_2 &\cdots &a^\prime_n \\
 a^\prime_2&a^\prime_3 &\cdots &a^\prime_{n+1} \\
\vdots &\vdots &\ddots &\vdots\\
 a^\prime_{n}&a^\prime_{n+1} &\cdots &a^\prime_{2n-1}
      \end{array}\right)\nonumber\\
&+&\psi\left[
\left(
\begin{array}{cccc}
 a_0&a_1 &\cdots &a_{n-1} \\
 a_1&a_2 &\cdots &a_{n} \\
\vdots &\vdots &\ddots &\vdots\\
 a_{n-1}&a_{n} &\cdots &a_{2n-2}
\end{array}\right)
\left(
\begin{array}{ccccc}
 0&a_0 &a_1 & \cdots & \cdots \\
 0& 0 &a_0 &a_1 & \cdots \\
 & &\ddots &\ddots &\vdots \\
 &\mbox{\O} &  &0 & a_0 \\
 &  &  &  & 0 
\end{array}\right)
\right.\nonumber\\
&+&
\left.\left(
\begin{array}{ccccc}
 0& & &  &\\
 a_0& 0 & &\mbox{\O}& \\
 a_1&a_0 & 0& & \\
 \vdots&\ddots & \ddots &\ddots&  \\
 \vdots& \cdots &a_1  & a_0&0 
\end{array}\right)
\left(
\begin{array}{cccc}
 a_0&a_1 &\cdots &a_{n-1} \\
 a_1&a_2 &\cdots &a_{n} \\
\vdots &\vdots &\ddots &\vdots\\
 a_{n-1}&a_{n} &\cdots &a_{2n-2}
\end{array}\right)
\right].\label{Toda:prod2}
\end{eqnarray}
\comment{ 
\begin{eqnarray}
&& \left(\begin{array}{cccc}
 a_1&a_2 &\cdots &a_n \\
 a_2&a_3 &\cdots &a_{n+1} \\
\vdots &\vdots &\ddots &\vdots\\
 a_{n}&a_{n+1} &\cdots &a_{2n-1}
      \end{array}\right)
=
 \left(\begin{array}{cccc}
 a^\prime_1&a^\prime_2 &\cdots &a^\prime_n \\
 a^\prime_2&a^\prime_3 &\cdots &a^\prime_{n+1} \\
\vdots &\vdots &\ddots &\vdots\\
 a^\prime_{n}&a^\prime_{n+1} &\cdots &a^\prime_{2n-1}
      \end{array}\right)\nonumber\\
&+&\psi\left[
\left(
\begin{array}{cccc}
 0&a_0^2 &\cdots &{\displaystyle \sum_{k=0}^{n-2}}a_ka_{n-2-k} \\
 0&a_0a_1 &\cdots &{\displaystyle\sum_{k=0}^{n-2}}a_ka_{n-1-k} \\
\vdots &\vdots &\ddots &\vdots\\
 0&a_{0}a_{n-1} &\cdots &{\displaystyle\sum_{k=0}^{n-2}}a_ka_{2n-3-k}\\
      \end{array}\right)\right.\nonumber\\
&+&\left.
\left(
\begin{array}{cccc}
 0&\cdots &0\\
 a_0^2&\cdots &a_{n-1} a_{0}\\
\vdots&\ddots &\vdots\\
{\displaystyle \sum_{k=0}^{n-2}}a_ka_{n-2-k} 
&\cdots &{\displaystyle \sum_{k=n-1}^{2n-3}}a_ka_{2n-3-k} \\
      \end{array}\right)
\right].\label{Toda:prod2}
\end{eqnarray}
} 
Applying scalar product of equation (\ref{Toda:prod2}) with
$(\Delta_{ij})$, we see that the first term of the right hand side of equation
(\ref{Toda:prod2}) gives $\tau_n^\prime$ and second and third terms give no 
contribution. Thus we have shown that equation (\ref{Toda:shift2})
holds. 

Equation (\ref{Toda:shift3}) is shown in a similar manner by considering the equality,
\begin{equation}
   D\left(\begin{array}{c}
   n\\
   n
	\end{array}\right)=
\left(\begin{array}{cccc}
 a_1&a_2 &\cdots &a_n \\
\vdots &\vdots &\ddots &\vdots\\
 a_{n-1}&a_{n} &\cdots &a_{2n-2}\\
 a_{n+1}&a_{n+2} &\cdots &a_{2n}
      \end{array}\right)\cdot
\left(\begin{array}{cccc}
 \Delta^\prime_{11}&\Delta^\prime_{12} &\cdots &\Delta^\prime_{1n} \\
 \Delta^\prime_{21}&\Delta^\prime_{22} &\cdots &\Delta^\prime_{2n} \\
\vdots &\vdots &\ddots &\vdots\\
 \Delta^\prime_{n1}&\Delta^\prime_{n2} &\cdots &\Delta^\prime_{nn} 
      \end{array}\right),\label{Toda:prod3}
\end{equation}
where $\Delta^\prime_{ij}$ denotes the $(i,j)$-cofactor of
$\tau_n^\prime= D\left(\begin{array}{c}
   n\\
   n+1
	\end{array}\right).$ \qed

The case $n<0$ of Theorem \ref{main1} is proved in a manner similar to the 
case of $n>0$, and the case $n=0$ is checked directly. Thus,
proof of Theorem \ref{main1} is completed.

Let us next prove Theorem \ref{main2}. Similarly to the proof of Theorem 
\ref{main1}, we concentrate on the case of $n>0$. We have the following
lemma:
\begin{lem}\label{disc:lem1}
 For $n\geq 1$, we have
\begin{equation}
 \varepsilon^{n-1}\kappa_n^{l+1}=\left|
\begin{array}{ccccc}
 c^l_0&c^l_1 &\cdots &c^l_{n-2}&C_0^{l+1} \\
 c^l_1&c^l_2 &\cdots &c^l_{n}&C_1^{l+1} \\
\vdots & \vdots &\ddots & \vdots&\vdots\\
 c^l_{n-1}&c^l_n &\cdots &c^l_{2n-2} &C_{n-1}^{l+1} 
      \end{array}\right|, \label{dToda:shift1}
\end{equation}
\begin{equation}
 \frac{\varepsilon^{2(n-1)}}{1-\varepsilon^2\Phi^{l+2}\Psi^l}\kappa_n^{l+2}
=
\left|
\begin{array}{ccccc}
 c^l_0&c^l_1 &\cdots &c^l_{n-2}&C_0^{l+1} \\
 c^l_1&c^l_2 &\cdots &c^l_{n}&C_1^{l+1} \\ \\
\vdots & \vdots &\ddots & \vdots&\vdots\\
 c^l_{n-2}&c^l_{n-1} &\cdots &c^l_{2n-3} &C_{n-2}^{l+1} \\
 C_0^{l+1}&C_1^{l+1} &\cdots &C_{n-2}^{l+1} &{\displaystyle 
\frac{C_0^{l+2}}{1-\varepsilon^2\Phi^{l+2}\Psi^l}}
      \end{array}\right|, \label{dToda:shift2}
\end{equation}
where $C_k^l$ is defined by
\begin{equation}
 C_k^l=c_k^l+\varepsilon\Psi^{l-2}\sum_{i=1}^kc_{k-i}^lc_{i-1}^{l-1}.\label{dToda:C}
\end{equation}
\end{lem}
Theorem \ref{main2} for the case of $n>0$ is derived as follows. We put 
\begin{equation}
 D\equiv  \frac{\varepsilon^{2n}}{1-\varepsilon^2\Phi^{l+2}\Psi^l}\kappa_{n+1}^{l+2}
=\left|
\begin{array}{ccccc}
 c^l_0&\cdots &c^l_{n-2}&c^l_{n-1}&C_0^{l+1} \\
 c^l_1&\cdots &c^l_{n-1}&c^l_{n}&C_1^{l+1} \\ \\
\vdots & \vdots &\ddots & \vdots&\vdots\\
 c^l_{n-2} &\cdots &c^l_{2n-4} &c^l_{2n-3}&C_{n-2}^{l+1} \\
 c^l_{n-1} &\cdots &c^l_{2n-3} &c^l_{2n-2}&C_{n-1}^{l+1} \\
 C_0^{l+1} &\cdots &C_{n-2}^{l+1} &C_{n-1}^{l+1}&{\displaystyle 
\frac{C_0^{l+2}}{1-\varepsilon^2\Phi^{l+2}\Psi^l}}
      \end{array}\right|. 
\end{equation}
Then we have from Lemma \ref{disc:lem1},
\begin{equation}
 D\left(\begin{array}{c}
   n\\
   n
\end{array}\right)= \frac{\varepsilon^{2(n-1)}}{1-\varepsilon^2\Phi^{l+2}\Psi^l}\kappa_n^{l+2},
\quad 
 D\left(\begin{array}{c}
   n+1\\
   n+1
\end{array}\right)=\kappa_n^l,
\end{equation}
\begin{equation}
  D\left(\begin{array}{c}
   n\\
   n+1
\end{array}\right)= D\left(\begin{array}{c}
   n+1\\
   n
\end{array}\right)= \varepsilon^{n-1}\kappa_n^{l+1},\quad
  D\left(\begin{array}{cc}
   n&n+1\\
   n&n+1
\end{array}\right)=\kappa_{n-1}^l.
\end{equation}
Therefore, Jacobi's identity (\ref{Toda:Jacobi}) yields
\begin{equation}
 \frac{\varepsilon^{2(n-1)}}{1-\varepsilon^2\Phi^{l+2}\Psi^l}\kappa_n^{l+2}~
\kappa_n^l - \left(\varepsilon^{n-1}\kappa_n^{l+1}\right)^2
=\frac{\varepsilon^{2n}}{1-\varepsilon^2\Phi^{l+2}\Psi^l}\kappa_{n+1}^{l+2}~
\kappa_{n-1}^l,
\end{equation}
which is equivlent to the discrete Toda equation (\ref{dToda:kappa}).

\noindent
{\it Proof of Lemma \ref{disc:lem1}:} We rewrite 
\[
 \kappa_n^{l+1}=
\left|\begin{array}{ccccc}
 c^{l+1}_0&c^{l+1}_1     &\cdots&c^{l+1}_{n-2} &c^{l+1}_{n-1}  \\
 c^{l+1}_1&c^{l+1}_2     &\cdots&c^{l+1}_{n-1}   &c^{l+1}_{n}    \\
\vdots & \vdots &\ddots &\vdots & \vdots\\
 c^{l+1}_{n-1}&c^{l+1}_n &\cdots&c^{l+1}_{2n-3}&c^{l+1}_{2n-2}  
      \end{array}\right|
\]
by using the recursion relation (\ref{dToda:rec1}) to obtain
eq. (\ref{dToda:shift1}).
We first add $j$-th column multiplied by
$\varepsilon\Psi^{l-1}c^{l}_{n-1-j}$ to $n$-th column for
$j=2,\cdots,n-1$. Next, adding $j$-th column multiplied by
$\Psi^{l-1}c^{l}_{n-2-j}$ to $n$-th column for
$j=1,\cdots,n-2$, we have from eq.(\ref{dToda:rec1}),
\[
 \kappa_n^{l+1}=
\left|\begin{array}{cccl}
 c^{l+1}_0     &\cdots&c^{l+1}_{n-2} &{\displaystyle -\frac{c^{l}_{n-2}}{\varepsilon}}  \\
 c^{l+1}_1     &\cdots&c^{l+1}_{n-1} &{\displaystyle -\frac{c^{l}_{n-1}}{\varepsilon}+\Psi^{l-1}(c_0^{l+1}-\varepsilon c_1^{l+1})c_{n-2}^l}    \\
\vdots & \vdots &\ddots &\vdots\\
 c^{l+1}_{n-2} &\cdots&c^{l+1}_{2n-4}&{\displaystyle -\frac{c^{l}_{2n-4}}{\varepsilon}+\Psi^{l-1}
\sum_{j=0}^{n-3}(c_j^{l+1}-\varepsilon c_{j+1}^{l+1})c_{2n-5-j}^{l}}\\
 c^{l+1}_{n-1} &\cdots&c^{l+1}_{2n-3}&{\displaystyle -\frac{c^{l}_{2n-3}}{\varepsilon}+\Psi^{l-1}
\sum_{j=0}^{n-2}(c_j^{l+1}-\varepsilon c_{j+1}^{l+1})c_{2n-4-j}^{l}}
      \end{array}\right|,
\]
Applying the similar procedure to $(n-1)$-th,$\cdots$,2nd columns, we obtain,
\begin{eqnarray*}
&& \kappa_n^{l+1}=\\
&&\left|\begin{array}{clcl}
 c^{l+1}_0 &{\displaystyle -\frac{c^{l}_{0}}{\varepsilon}} &\cdots &{\displaystyle -\frac{c^{l}_{n-2}}{\varepsilon}}  \\
 c^{l+1}_1 &{\displaystyle -\frac{c^{l}_{1}}{\varepsilon}
+\Psi^{l-1}(c_0^{l+1}-\varepsilon c_1^{l+1})c_{0}^l} &\cdots&
{\displaystyle -\frac{c^{l}_{n-1}}{\varepsilon}
+\Psi^{l-1}(c_0^{l+1}-\varepsilon c_1^{l+1})c_{n-2}^l}    \\
\vdots & \vdots &\ddots &\vdots\\
 c^{l+1}_{n-2} &{\displaystyle -\frac{c^{l}_{n-2}}{\varepsilon}+\Psi^{l-1}
\sum_{j=0}^{n-3}(c_j^{l+1}-\varepsilon c_{j+1}^{l+1})c_{n-3-j}^{l}}
&\cdots&{\displaystyle -\frac{c^{l}_{2n-4}}{\varepsilon}+\Psi^{l-1}
\sum_{j=0}^{n-3}(c_j^{l+1}-\varepsilon c_{j+1}^{l+1})c_{2n-5-j}^{l}}\\
 c^{l+1}_{n-1} &{\displaystyle -\frac{c^{l}_{2n-3}}{\varepsilon}+\Psi^{l-1}
\sum_{j=0}^{n-2}(c_j^{l+1}-\varepsilon c_{j+1}^{l+1})c_{n-2-j}^{l}}
&\cdots&{\displaystyle -\frac{c^{l}_{n-1}}{\varepsilon}+\Psi^{l-1}
\sum_{j=0}^{n-2}(c_j^{l+1}-\varepsilon c_{j+1}^{l+1})c_{2n-4-j}^{l}}
      \end{array}\right|,
\end{eqnarray*}
Next, we apply the similar procedure in vertical direction. For
$k=2,\cdots n$, we add $j$-th row multiplied by
$\varepsilon\Psi^{l-1}(c_{j-1}^{l+1}-c_{j}^{l+1})$ to $k$-th row for
$j=1,\cdots,k-1$. Then we get
\begin{eqnarray*}
&&\kappa_n^{l+1}=
\left|\begin{array}{cccc}
 C_0^{l+1}    &-\frac{c^{l}_0}{\varepsilon} &\cdots &-\frac{c^{l}_{n-1}}{\varepsilon} \\
 C_1^{l+1}    &-\frac{c^{l}_1}{\varepsilon} &\cdots &-\frac{c^{l}_n}{\varepsilon} \\
\vdots        & \vdots &\ddots & \vdots\\
 C_{n-1}^{l+1}&-\frac{c^{l}_{n-1}}{\varepsilon}&\cdots &-\frac{c^{l}_{2n-2}}{\varepsilon}
      \end{array}\right|\\
&=&\varepsilon^{-(n-1)}\left|\begin{array}{cccc}
c^{l}_0    &\cdots &c^{l}_{n-1} & C_0^{l+1}\\
c^{l}_1    &\cdots &c^{l}_n     & C_1^{l+1}\\
 \vdots    &\ddots & \vdots     & \vdots\\
c^{l}_{n-1}&\cdots &c^{l}_{2n-2}& C_{n-1}^{l+1}
      \end{array}\right|,
\end{eqnarray*}
where $C_k^{l+1}$ is defined recursively by
\begin{equation}
 C_k^{l+1}=c_k^{l+1}+\sum_{j=0}^{k-1}\varepsilon\Psi^{l-1}(c_{k-1-j}^{l+1}
-\varepsilon c_{k-j}^{l+1})C_j^{l+1},\quad C_0^{l+1}=c_0^{l+1},
\end{equation}
We can verify that $C_k^{l}$ is actually given as eq.(\ref{dToda:C}) by
induction. Thus we have proved eq.(\ref{dToda:shift1}). Since
eq.(\ref{dToda:shift2}) is proved by similar calculation, we 
omit the detail. This completes the proof of Lemma
\ref{disc:lem1}.\qed\\

The above discussion proves Theorem \ref{main2} for the case of
$n>0$. The case of $n<0$ is proved similarily, and the case of $n=0$ is
checked immediately. Thus we have proved Theorem \ref{main2}.

\section{Application for Painlv\'e equations}

It is established by K.~Okamoto that the
$\tau$ finctions of the Painlev\'e equations $P_{II}, \ldots, P_{VI}$
satisfy the Toda equation.
Recall that 
each of the Painlev\'e equations $P_J$ ($J=II,\ldots,VI$)  
can be written as 
a Hamiltonian system 
\begin{equation}
\delta q={\partial H \over \partial p}, \quad
\delta p=-{\partial H \over \partial q},
\end{equation}
where $H$ is a certain polynomial in $q$, $p$, and 
the derivation $\delta$ is given by 
$\delta=\partial_t$ for $P_{II}, P_{IV}$,
$\delta=t \partial_t$ for $P_{III}, P_{V}$,
and $\delta=t(t-1) \partial_t$ for $P_{VI}$,
respectively. 
(The explicit formula for $H$ will be given below.)
The $\tau$ function, which we denote $\sigma$, 
is defined up to constant factor as
\begin{equation}
\delta(\log \sigma)=H.
\end{equation}
By using a B\"acklund transformation (Schlesinger transformation)
$T$ which acts as a translation on the parameter space,
one can define a sequence of functions $\sigma_n$ by 
choosing an appropriate normalization factor for $T^n(\sigma)$ 
($n\in \mathbb{Z}$). 

\begin{theorem} (Okamoto \cite{O1}) 
The sequence of $\tau$ functions  $\sigma_n$ for the 
Painlev\'e equations $P_{J}$ satisfies the Toda equation 
\begin{equation}
(\delta^2 \sigma_n) \sigma_n-(\delta \sigma_n)^2
=\sigma_{n+1} \sigma_{n-1}.\label{eq:s.toda}
\end{equation}
\end{theorem}
We will review the derivation of the Toda equation in the 
next section for completeness.
Let us put 
\begin{equation}
\sigma_n=\sigma_0 ({\sigma_1 \over \sigma_0})^n T_n,
\end{equation}
then the function $T_n$ satisfy the recurrence relation
\begin{equation}
T_{n+1} T_{n-1}=
(\delta^2 T_n) T_n-(\delta T_n)^2+(n \delta v+u) {T_n}^2,
\label{eq:Trec}
\end{equation}
with initial condition $T_0=T_1=1$.
Here $u=\delta^2 \log \sigma_0$, 
$v=\delta \log {\sigma_1 \over \sigma_0}$. 
The following has been conjectured by H.~Umemura.
(see \cite{Um2} for example)

\begin{theorem}
Let $(q,p)$ be a solution for $P_{J}$ and define
functions $T_n=T_n(q,p,t)$ through the recurrence relation
(\ref{eq:Trec}). 
Then the function $T_n$ is polynomial in $q,p$. 
\end{theorem}

\noindent {\it Proof}.
We will show for $n \geq 0$.
The case $n \leq 0$ is similar.
Note that the functions $\tau_n=\sigma_n/\sigma_0$ ($n\in \mathbb{Z}$) 
satisfy the equation
\begin{equation}
 \tau_n^{\prime\prime}\tau_n - \tau_n^{\prime 2}=\tau_{n+1}\tau_{n-1}-\psi\varphi\tau_n^2,
\end{equation}
with 
\begin{equation}
\psi=\frac{\sigma_{-1}}{\sigma_0},\quad \quad \varphi=\frac{\sigma_1}{\sigma_0}.
\end{equation}
{}
From the determinant formula (Theorem \ref{main1}),
the solution of the Toda equation (\ref{eq:s.toda}) is given by
\begin{equation}
\sigma_n=\sigma_0\tau_n=
\sigma_0 \det(a_{i+j})_{0 \leq i,j \leq n-1}.
\end{equation}
We introduce  $g_i$ ($i\in\mathbb{N})$ by setting 
$a_i=\sigma_1g_i/\sigma_0=\varphi g_i$, so that 
\begin{equation}
\sigma_n=
\sigma_0 ({\sigma_1 \over \sigma_0})^n 
\det(g_{i+j})_{0 \leq i,j \leq n-1}.
\end{equation}
Putting $u=\delta^2 \log \sigma_0=\psi \varphi$ and
$v=\delta \log \varphi=\delta \log {\sigma_1 \over \sigma_0}$,
we can rewrite the recurrence formula for $a_n$ to that of $g_n$:  
\begin{equation}
g_0=1, \quad {\rm and} \quad
g_n=\delta g_{n-1}+v g_{n-1}+u 
\sum_{i+j=n-2\atop i,j\geq 0} g_i g_j, \quad (n \geq 1).
\end{equation}
For the Painlev\'e $\tau$ functions,
$u$, $v$ (these are essentially $H$ and $Y$ given in the next section)
and their $\delta$-derivatives
are all polynomials in $q,p$.
Hence the coefficients $g_k$ and their determinants $T_n$ are
also polynomials in $q,p$.\qed

\def\aa{\alpha}

\section{Derivation of the Toda equations}

In this section we will review the derivation of the Toda
equation following the work by Okamoto \cite{O1,O2,O3,O4}.

\subsection{The second Painlev\'e equation: $P_{II}$}
The Hamiltonian is
\begin{equation}
H={1 \over 2}p^2-(q^2+{1 \over 2} t) p-\aa_1 q.
\end{equation}
The equation for $y=q$ is the Painlev\'e equation $P_{II}$
given as follows:
\begin{equation}
y''=2 y^3+t y+a,
\end{equation}
with $a=\alpha_1-{1 \over 2}$.

The B\"acklund transformations are given by
\begin{equation}
\begin{array}{|c||c|c|c|c|}
\hline
x & \aa_0 & \aa_1 & p & q  \\
\hline
s_0(x) & -\aa_0 & \aa_1 +2 \aa_0
& p+{4 \aa_0 q \over f}+{2 {\aa_0}^2 \over f^2} & q+{\aa_0 \over f}  \\
s_1(x) & \aa_0+2 \aa_1 & -\aa_1 & p & q+{\aa_1 \over p}  \\
\pi(x) & \aa_1 & \aa_0 & -f & -q  \\
\hline
\end{array}
\end{equation}
where $\aa_0=1-\aa_1$ and $f=p-2 q^2-t$.

Let $T=\pi s_1$ be the translation which acts on the parameter as
$T(\aa_0,\aa_1)=(\aa_0+1,\aa_1-1)$, then we have 
\begin{equation}
T(H)-H=Y=q, \quad
\partial_t(H)=-{p \over 2},
\end{equation}
and
\begin{equation}
\partial_t(\log \partial_t H)=T(H)-2 H+T^{-1}(H).
\end{equation}
Hence we have the Toda equation :
\begin{equation}
{\partial_t}^2 \log \sigma_n=
c(n) {\sigma_{n-1} \sigma_{n+1} \over {\sigma_{n}}^2},
\end{equation}
where $c(n)$ is a non-zero constant.
This can be transformed to the standard Toda equation
by changing the normalization as
\begin{equation}
\sigma_n \rightarrow C_n\, \sigma_n.
\end{equation}

\subsection{The third Painlev\'e equation : $P_{III}$} 
The Hamiltonian is
\begin{equation}
H_{III}=q^2 p^2-(q^2+v_1 q-t)p-{1 \over 2}(v_1+v_2) q.
\end{equation}
The equation for $y=q/s$ ($t=s^2$) is given by
the third Painleve equation $P_{III}$
\begin{equation}
{d^2 y \over ds^2}={1 \over y}({dy \over ds})^2-{1 \over s}{dy \over ds}+
+{1 \over s}(a y^2+b)+c y^3+{d \over y},
\end{equation}
with
\begin{equation}
a=-4 v_2, \quad
b=4(v_1+1), \quad
c=4, \quad
d=-4.
\end{equation}

The B\"acklund transformations are
\begin{equation}
\begin{array}{|c||c|c|c|c|c|}
\hline
x & v_1 & v_2 & p & q & t \\
\hline
s_0(x) & -1-v_2 & -1-v_1
& {q \over t}(q(p-1)-{1 \over 2}(v_1-v_2))+1
& -{t \over q} & t\\
s_1(x) & v_2 & v_1 & p & q+{{1 \over 2}(v_2-v_1) \over p-1} & t \\
s_2(x) & v_1 & -v_2 & 1-p & -q  & -t\\
\hline
\end{array}
\end{equation}

Let $T=s_0 s_2 s_1 s_2$ be the translation
which acts on the parameter as
$T(v_1,v_2)=(v_1+1,v_2+1)$,
then 
\begin{equation}
T(H)-H=Y=q(1-p), \quad
\partial_t(H)=p.
\end{equation}
and
\begin{equation}
\delta \log (\partial_t H)=T(H)-2 H+T^{-1}(H).
\end{equation}
Hence we have the Toda equation 
\begin{equation}
{\partial_t}\delta \log \sigma_n 
=c(n){\sigma_{n-1} \sigma_{n+1} \over {\sigma_{n}}^2},
\end{equation}
where $c(n)$ is a non-zero constant.
This equation can be transformed to the standard Toda equation by
the change of the normalization as
\begin{equation}
\sigma_n \rightarrow C_n\, t^{n^2/2} \sigma_n.
\end{equation}

\subsection{The fourth Painlev\'e equation : $P_{IV}$}
The Hamiltonian is
\begin{equation}
H_{IV}=(p-q-2 t) p q-2 \aa_1 p-2 \aa_2 q.
\end{equation}
Equation for $y=q$ is given by
\begin{equation}
y''={{y'}^2 \over 2y}+{3 y^3 \over 2}+4 t y^2
+2 (t^2 -a) y+{b \over y},
\end{equation}
with
\begin{equation}
a=\aa_0-\aa_2, \quad
b=-2 {\aa_1}^2.
\end{equation}
The B\"acklund transformations are 
\begin{equation}
\begin{array}{|c||c|c|c|c|c|}
\hline
x & \aa_0 &\aa_1 & \aa_2 & p & q \\
\hline
s_0(x) & -\aa_0 & \aa_1+\aa_0 & \aa_2+\aa_0
& p+{2 \aa_0 \over f} & q+{2 \aa_0 \over f} \\
s_1(x) & \aa_0+\aa_1 & -\aa_1 & \aa_2+\aa_1 
& p-{2 \aa_1 \over q} & q \\
s_2(x) & \aa_0+\aa_2 & \aa_1+\aa_2 & -\aa_2 
& p & q+{2 \aa_2 \over p} \\
\pi(x) & \aa_1 & \aa_2 & \aa_0
& -f & -p \\
\hline
\end{array}
\end{equation}
where $\aa_0=1-\aa_1-\aa_2$ and $f=p-q-2 t$.

Let $T=\pi s_2 s_1$ be the translation
which acts on the parameter as
$T(\aa_0,\aa_1,\aa_2)=(\aa_0+1,\aa_1-1,\aa_2)$,
then 
\begin{equation}
T(H)-H=Y=2 p, \quad
\partial_t(H)=-2 p q.
\end{equation}
and
\begin{equation}
\delta \log (\partial_t H+4 \aa_1)=T(H)-2 H+T^{-1}(H).
\end{equation}
Hence we have the Toda equation 
\begin{equation}
{\partial_t}\delta \log \sigma_n +4 (\aa_1-n)
=c(n){\sigma_{n-1} \sigma_{n+1} \over {\sigma_{n}}^2},
\end{equation}
where $c(n)$ is a non-zero constant.
The change of normalization is given by
\begin{equation}
\sigma_n \rightarrow C_n\, e^{2(\aa_1-n) t^2} \sigma_n.
\end{equation}

\subsection{The fifth Painlev\'e equation : $P_{V}$}
The Hamiltonian is
\begin{equation}
H_{V}=p(p+t)q (q-1)+\aa_2 q t-\aa_3 p q-\aa_1 p(q-1).
\end{equation}
Equation for $y=1-1/q$ is given by
\begin{equation}
y''=({1 \over 2y}+{1 \over y-1}) (y')^2-{y' \over t}+
{(y-1)^2  \over t^2}(a y+{b \over y})+c {y \over t}+
d {y(y+1) \over y-1},
\end{equation}
where 
\begin{equation}
a={{\aa_1}^2 \over 2}, \quad
b=-{{\aa_3}^2 \over 2}, \quad
c=\aa_0-\aa_2, \quad
d=-{1 \over 2}.
\end{equation}
The B\"acklund transformations are 
\begin{equation}
\begin{array}{|c||c|c|c|c|c|c|}
\hline
x & \aa_0 &\aa_1 & \aa_2 & \aa_3 & p & q \\
\hline
s_0(x) & -\aa_0 & \aa_1+\aa_0 & \aa_2 & \aa_3+\aa_0
& p & q+{\aa_0 \over p+t} \\
s_1(x) & \aa_0+\aa_1 & -\aa_1 & \aa_2+\aa_1 & \aa_3 
& p-{\aa_1 \over q} & q \\
s_2(x) & \aa_0  & \aa_1+\aa_2 & -\aa_2 & \aa_3+\aa_2
& p & q+{\aa_2 \over p} \\
s_3(x) & \aa_0+\aa_3 & \aa_1 &\aa_2+\aa_3 & -\aa_3
& p+{\aa_3 \over q-1} & p \\
\pi(x) & \aa_1 & \aa_2 & \aa_3 & \aa_0
& t(q-1) & -{p \over t} \\
\hline
\end{array}
\end{equation}
where $\aa_0=1-\aa_1-\aa_2-\aa_3$.

Let $T=\pi s_3 s_2 s_1$ be the translation
which acts on the parametrs as
$T(\aa_0,\ldots,\aa_3)=(\aa_0+1,\aa_1-1,\aa_2,\aa_3)$,
then 
\begin{equation}
T(H)-H=Y=p(q-1), \quad
\partial_t(H)=p q(q-1)+\aa_2 q,
\end{equation}
and
\begin{equation}
\delta \log (\partial H+\aa_1)=
T(H)-2 H+T^{-1}(H).
\end{equation}
Hence, we obtain the Toda equation 
\begin{equation}
{\partial_t}{\delta} \log \sigma_n+(\aa_1-n)=
c(n){\sigma_{n-1} \sigma_{n+1} \over 
{\sigma_{n}}^2},
\end{equation}
where $c(n)$ is a non-zero constant.
The change of the normalization is given by
\begin{equation}
\sigma_n \rightarrow C_n\, t^{n^2/2} e^{(\aa_1-n)t} \sigma_n.
\end{equation}

\subsection{The sixth Painlev\'e equation $P_{VI}$}
The Hamiltonian is
\begin{equation}
H=q(q-1)(q-t)p^2-[\alpha_4(q-1)(q-t)+\alpha_3 q(q-t)+(\alpha_0-1)q(q-1)]p+
\alpha_2(\alpha_1+\alpha_2)(q-t).
\end{equation}
The equation for $y=q$ is given by
\begin{eqnarray*}
&&y''={1 \over 2}\big({1 \over y}+{1 \over y-1}+
{1 \over y-t}\big){y'}^2-\big({1 \over t}+{1
\over t-1}+{1 \over y-t}\big)y'\\
&&+{y(y-1)(y-t) \over t^2(t-1)^2}
\big[a+b{t \over y^2}+
c{t-1 \over (y-1)^2}+
d{t(t-1) \over (y-t)^2}\big], 
\end{eqnarray*}
with
\begin{equation}
a={\alpha_1}^2/2, \quad
b=-{\alpha_4}^2/2, \quad
c={\alpha_3}^2/2, \quad
d=-({\alpha_0}^2-1)/2.
\end{equation}
($\alpha_0+\alpha_1+2 \alpha_2+\alpha_3+\alpha_4=1$).
%
%
The B\"acklund transformations are as follows
\begin{equation}
\begin{array}{|c||c|c|c|c|c|c|c|}
\hline
x & \aa_0 & \aa_1 & \aa_2 & \aa_3 & \aa_4 & p & q  \\
\hline
s_0(x) & -\aa_0 & \aa_1 & \aa_2+\aa_0 & \aa_3 & \aa_4 
& p-{\aa_0 \over q-t} & q  \\
s_1(x) & \aa_0 & -\aa_1 & \aa_2+\aa_1 & \aa_3 & \aa_4 
& p & q  \\
s_2(x) & \aa_0+\aa_2 & \aa_1+\aa_2 & -\aa_2 & \aa_3+\aa_2 & \aa_4+\aa_2 
& p & q \rightarrow q+{\alpha_2 \over p}  \\
s_3(x) & \aa_0 & \aa_1 & \aa_2+\aa_3 & -\aa_3 & \aa_4 
& p \rightarrow p -{\alpha_3 \over q-1} & q  \\
s_4(x) & \aa_0 & \aa_1 & \aa_2+\aa_4 & \aa_3 & -\aa_4 
& p \rightarrow p-{\alpha_4 \over q} & q  \\
\hline
r(x) & \aa_1 & \aa_0 & \aa_2 & \aa_4 & \aa_3 
& -{p(q-t)^2+\alpha_2(q-t) \over t(t-1)} &  
q \rightarrow {(q-1)t \over (q-t)} \\
\hline
\end{array}
\end{equation}

Let $T=s_0s_2s_3s_4s_2s_0 r$ be the translation
which acts on the parameters as 
$T(\aa_0,\ldots,\aa_4)=
(\aa_0-1,\aa_1+1,\aa_2,\aa_3,\aa_4)$,
then we have
\begin{equation}
T(H)-H=Y=p q(1-q)+\alpha_2(t-q), \quad
\partial_t H=p^2 q (1-q)-\aa_2 (\aa_1+\aa_2)+(\aa_3+\aa_4) pq-\aa_4 p,
\end{equation}
and
\begin{equation}
\delta \log(\partial_t H+(\aa_1+\aa_2)(1-\aa_0))=
T(H)-2 H+T^{-1}(H).
\end{equation}
The Toda equation is given as 
\begin{equation}
{\partial_t}{\delta} \log \sigma_n+(\aa_1+\aa_2+n)(1-\aa_0+n)=
c(n) {\sigma_{n-1} \sigma_{n+1} \over 
{\sigma_{n}}^2},
\end{equation}
where $c(n)$ is a non-zero constant.
The change of the normalization is given by
\begin{equation}
\sigma_n \rightarrow C_n\, (t(t-1))^{(\aa_1+\aa_2+n)(1-\aa_0+n)} \sigma_n.
\end{equation}

\section{Conclusion}
In this paper, we have presented determinant formulas for the general
solutions of both Toda and discrete Toda equations. We have applied the
results to the Painlev\'e equations, and obtained a new direct proof for 
the polynomiality of the solutions of Painlev\'e equations.
\vskip10mm
\noindent
{\bf {Acknowledgment.}}
We wish to thank Professor H.~Umemura for stimulating discussions.

\end{document}